\documentclass[runningheads]{llncs}

\usepackage[T1]{fontenc}
\usepackage{graphicx}
\usepackage{glossaries}
\usepackage[hidelinks]{hyperref}
\usepackage{cleveref}
\usepackage{microtype}
\usepackage{tikz}
\usepackage{enumitem}
\usepackage{mdframed}
\usepackage[backend=biber,
  style=lncs,
  doi=false,
  isbn=false
]{biblatex}
\usepackage{csquotes}
\usepackage{xcolor}

\usepackage[untagged,highstructure]{accessibility}

\newacronym{vcs}{VCS}{version control system}
\newacronym{cla}{CLA}{contributor license agreement}
\newacronym{dao}{DAO}{decentral autonomous organization}
\newacronym{otf}{OTF}{Open Technology Fund}
\newacronym{osi}{OSI}{Open Source Initiative}
\newacronym{stf}{STF}{Sovereign Tech Fund}
\newacronym{sta}{STA}{Sovereign Tech Agency}
\newacronym{foss}{FOSS}{free and open source software}
\newacronym{ostif}{OSTIF}{Open Source Technology Improvement Fund}
\newacronym{openssf}{OpenSSF}{Open Secure Software Foundation}

\usepackage{orcidlink}
\renewcommand{\orcidID}[1]{\orcidlink{#1}}

\begin{document}
\title{How Reliable Are FOSS Popularity Metrics? Analyzing the Effort Required for Spoofing Common Software Popularity Metrics}
\titlerunning{Reliability of FOSS Popularity Metrics}
%
\author{Ben Swierzy\inst{1,2}\orcidID{0009-0003-0485-4791} \and
    Timo Pohl\inst{1}\orcidID{0009-0002-3760-7976} \and
    Marc Ohm\inst{1,2}\orcidID{0000-0002-2913-5270} \and
    Michael Meier\inst{1,2}\orcidID{0009-0006-8199-5004}}

\authorrunning{B. Swierzy et al.}
\institute{University of Bonn, Germany \\
    \email{\{swierzy,pohl,ohm,mm\}@cs.uni-bonn.de} \and
    Fraunhofer FKIE, Germany}

\maketitle              
\begin{abstract}
  Quantitative metrics derived from software repositories and package ecosystems are widely used to assess the impact, popularity, maintenance, and criticality of free and open source software (FOSS) projects.
  However, these metrics are often assumed to be reliable despite their potential susceptibility to manipulation.
  Prior empirical software engineering and security research deployed these in a variety of ways which assume they indeed capture project impact and popularity.
  Yet, the extent to which these underlying signals can be spoofed in practice, and the consequences this has for downstream uses of the metrics, has received little focused attention.
  To address this gap, the paper decomposes existing combined metrics into atomic metric categories, analyzes their spoofing effort under a maintainer-centered threat model, and investigates a real-world sybil attack on npm connected to an impact-based reward mechanism.
  The analysis finds that many metric categories, especially commit data, issue-tracker activity, downloads, repository contents, and dependency relations, are manipulable with low to moderate effort, and it identifies a sybil attack comprising more than 70,000 spam packages on npm.
  These results imply that quantitative FOSS metrics should be used with much greater caution in software engineering research and practice, particularly for ranking, dataset construction, and any allocation or evaluation process that turns metrics into optimization targets.

\end{abstract}
\textit{Note: This paper is a shortened version of \fullcite{self}.. A pre-print of the full version is available at \url{https://arxiv.org/abs/2505.05897v1}}
\section{Introduction}

During the past decade, \gls{foss} has established itself as essential in modern digital systems.
It is recognized as digital infrastructure, building the foundation for new developments.
\Gls{foss} is viewed as a digital common good beneficial for society.
While individuals are able to rapidly bootstrap new projects, developing products has become drastically cheaper for the industry as well.
Overall, digital innovation is thriving.~\cite{Eghbal2016}

Within this large body of \gls{foss}, there naturally is a gradient with respect to the quality, impact and criticality, which typically manifests in a project's popularity.
All of these aspects are supposedly captured in a large variety of project metrics, which aid in various use cases.
For practitioners, these metrics assist in selecting new project dependencies if multiple options exist~\cite{LariosVargas2020}.
For end users, such metrics represent an extrinsic factor for trust in software products~\cite{Hou2023}.
For funding programs, these metrics may hint what projects have a larger societal impact and better fit the objective of the fund.
Given this significance of popularity metrics, the impact of spoofing is a relevant threat.
Therefore, we investigate the following research question.
\begin{enumerate}[label=\bfseries RQ\arabic*\normalfont,left=2mm]
    \item How reliable are common software popularity metrics as typically used for the assessment of \gls{foss} projects?
\end{enumerate}
Our contributions are three-fold.
We compile practical examples for automatic manipulation of impact metrics, uncover a sybil attack on npm which could affect future research building on its registry data, and propose strategies how to deal with unreliable metrics.

The remainder of the paper is structured as follows.
\Cref{sec:metric-reliability} compiles common popularity metrics, collects evidence for spoofing attempts and compares the associated effort.
\Cref{sec:resilience} discusses ways forward and strategies how to handle the impact of unreliable metrics.
Related work is presented in \Cref{sec:related-work}.
We conclude in \Cref{sec:conclusion}.

\section{Reliability of Popularity Metrics for Software}
\label{sec:metric-reliability}

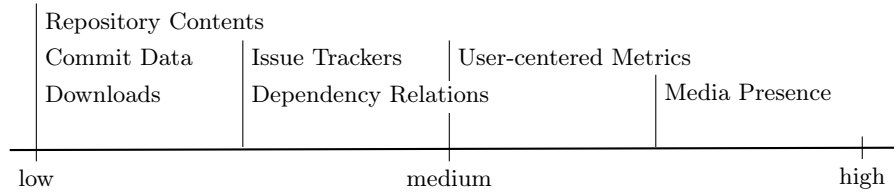
\begin{figure}
  \alt{Horizontal scale rating the effort to spoof atomic metrics. Categories are low (repository contents, commit data, downloads), low to medium (issue trackers, dependency relations), medium (user-centered metrics), medium to high (media presence), and high (none).}
  \centering
  \setlength\fboxsep{2pt}
  \begin{tikzpicture}
    \node (low) at (0,0) {low};
    \node (lowToMedium) at (.225\textwidth,0) {};
    \node (medium) at (.45\textwidth,0) {medium};
    \node (mediumToHigh) at (.675\textwidth,0) {};
    \node (high) at (.9\textwidth,0) {high};
    
    \draw[thick] ([yshift=0.15cm]low.north west) -- ([yshift=0.15cm]high.north east);
    \draw (high.north) -- ++(0,.3);
    
    \node[anchor=south west, text width=4cm] (mediumToHighLabel) at ([yshift=0.4cm]mediumToHigh) {\baselineskip=15pt{Media Presence}\\~\par};
    \draw (mediumToHighLabel.north west) -- ([yshift=0.3cm]mediumToHigh.north);
    \node[anchor=south west, text width=4cm] (mediumLabel) at ([yshift=0.4cm]medium) {\baselineskip=14pt{User-centered Metrics}\\~\\~\par};
    \draw (mediumLabel.north west) -- (medium.north);
    \node[anchor=south west, text width=4cm] (lowToMediumLabel) at ([yshift=0.4cm]lowToMedium) {\baselineskip=14pt{Issue Trackers}\\\colorbox{white}{\hspace*{-2pt}Dependency Relations}\\~\par};
    \draw (lowToMediumLabel.north west) -- ([yshift=0.3cm]lowToMedium.north);
    \node[anchor=south west, text width=4cm] (lowLabel) at ([yshift=0.4cm]low) {\baselineskip=14pt{Repository Contents}\\Commit Data\\Downloads\\~\par};
    \draw (lowLabel.north west) -- (low.north);
    
  \end{tikzpicture}
  \caption{Spoofing effort for categories of atomic metrics}
  \label{fig:effort-line}
\end{figure}

Rating the criticality of a project and measuring its impact is an essential step in many scenarios.
It is important to consider the reliability of impact and popularity metrics for FOSS to be able to assess the likelihood of the presented fraud scenarios.
In this section, we analyze 4 existing combined metrics, namely \gls{openssf}'s criticality score~\cite{Arya2023}, npm's quality/popularity/maintenance scores, teaRank~\cite{tearank2025}, and CHAOSS's project popularity metric~\cite{CHAOSS2024}, by dissecting them into their components, referred to as \emph{atomic metrics}.
To deal with subtle differences between some atomic metrics, we categorize them by their underlying data source.
All categories are visualized in~\Cref{fig:effort-line}.
We discuss their reliability with respect to a project maintainer with write access to a project's repository contents and its presence on social coding platforms.

\subsubsection{Commit Data}

A commit is the essential unit of a \gls{vcs}.
Active project maintenance implies regular commits.
Therefore, atomic metrics derived from commit timestamps are commonly used.
However, this type of data is highly unreliable and easily spoofed.
Social coding platforms display commit data unaltered.
This allows maintainers to create an arbitrary history and push that onto the platform to fulfill all desired metrics.
While commits may be signed to verify the identity of the commit author, it does not help if the attacker is a project maintainer.
There are tools exploiting this behavior, for example, the \texttt{github-activity-generator} creating a commit history to obtain a custom activity graph in GitHub.
It should be noted that the tool explicitly discourages its use to \textquote{misrepresent professional contributions or coding activity}~\cite{Shpota2025}.

\subsubsection{Issue Trackers}

Many \gls{foss} projects facilitate the issue trackers provided on social coding platforms.
While lots of open issues do not imply meaningful statements, an active usage of the issue tracker suggests interest and maintenance of a project.
The lifetime of issues offers multiple atomic features to be extracted.
Among the associated metrics, the ratio of open to closed issues and the average duration until an issue is closed are prime examples.
All popular issue trackers offer APIs for automatic interactions and integration of external clients.
A plethora of well-known large projects utilizes bots to assist with the issue management.
Examples are TensorFlow\footnote{\url{https://github.com/tensorflow/tensorflow}} which automatically assigns developers, and Go\footnote{\url{https://github.com/golang/go}} which has an AI assistant automatically tagging the issue and searching links for related information.
React\footnote{\url{https://github.com/facebook/react}} utilizes an officially provided GitHub Actions workflow 
to automatically close stale open issues.
While we do not assume any malicious motives in this case, this directly boosts an atomic issue metric determining the maintenance score on npm.
Other platform features such as publishing explicit releases offer comparable data for atomic metrics.
For maintainers, these are as easily spoofable as issue-related data due to public APIs.
Therefore, legitimate tools like \texttt{semantic-release}\footnote{\url{https://github.com/semantic-release/semantic-release}} increase the score in associated metrics.

\subsubsection{User-centered Metrics}

Atomic metrics are considered to be user-centered if every user can increase this metric at most by one.
They are platform-specific with examples being stars, forks, subscriptions or contributors.
Of all categories with implemented measures, user-centered atomic metrics are the most difficult metric to spoof, requiring the attacker to automate account creation and violating the terms of service.
Nonetheless, this can be outsourced to paid services.
A report~\cite{Eldeeb2023} describes stars to be successfully purchasable for 0.08 EUR each.
However, the automatically created accounts for these are detected and removed along with the stars within a month.
In contrast, \textquote{quality} stars cost 0.8 EUR each and are backed by accounts not as prone to bot detection.
Services to buy forks and subscribers are found as easily.
Therefore, given enough monetary or technical resources it seems to be possible to spoof user-centered metrics.

\subsubsection{Downloads}

Impactful projects are utilized by many users and, thus, deployed on many systems.
Analyzing recent download counts is a proxy metric suggesting to approximate the deployment count.
However, there is no clear connection and a widely deployed project does not need to have many recent downloads.
More severely, the process of downloading is easy to automate.
For npm, there even exist tools such as \texttt{npm-increaser-downloads}\footnote{\url{https://github.com/MinhOmega/npm-increaser-downloads}} offering an optimized implementation.
Though it considerably wastes resources of the package registry, we did not find information on restricted counting, e.g., only once per IP address and day.
While this does not defend against such an attack, it significantly increases the effort required by an attacker with only minor overhead for the registry.

\subsubsection{Repository Contents}

The quality metric of npm solely focuses on atomic metrics derived from the contents of a repository such as the existence of a license or the use of linters.
The repository contents are trivial to adjust for a maintainer and, thus, it requires low effort to achieve a high score in this metric.
Though in this case, we refrain from classifying such adjustments as spoofing, since no information is illegitimately represented.

\subsubsection{Media Presence}

CHAOSS' project popularity metric~\cite{CHAOSS2024} enhances its score by incorporating data from sources beyond social coding platforms and package registries.
The atomic metrics range from social media mentions over job postings requesting project skills to event participation.
Most are challenging to determine automatically and as such, their applicability is likely limited to the largest \gls{foss} projects.
Accordingly, these atomic metrics are difficult to spoof as their methods of measurement are not well specified.

\subsubsection{Dependency Relations}
\label{sec:dependency-relations}

Packages support an effective development process by providing functionality in a re-usable manner.
Besides customer-facing end products, this also backs the development of new packages, resulting in a dependency network.
This offers a unique view on the impact of projects and is integrated into all considered impact metrics.
More specific, either the number of dependent projects or the more complicated and holistic view of teaRank, an algorithm inspired by page rank that is designed to provide a passive source of income for impactful \gls{foss}~\cite{tearank2025}, are employed.
These metrics can be spoofed by creating bogus packages, referencing the maintainer's project either directly or transitively as dependency.
For uncurated package repositories, we deem this attack to require low to medium effort, depending on the efficacy of its automatic spam detection mechanisms.
Since we were unable to find existing indicators for the practical relevance of this issue, we subsequently uncover and analyze an incident on npm to provide evidence for this assessment.

\paragraph{Sybil Attacks on npm}

A passive rewarding system such as teaRank has the potential to incentivize maintainers to enhance their impact through dishonest methods.
Its technical description~\cite{tearank2025} acknowledges two types of attacks.
Width attacks introduce lots of dependents pointing to a single package.
Tree attacks create long dependency chains.
It is stated that both attacks are prevented by tracking the width and tree limit of a package and flagging it as potential spam if (secret) thresholds are surpassed.
During our work, we manually inspected packages on the tea testnet and found the majority of the projects denoted as most impactful, i.e., having the highest teaRanks, to have either been unpublished or replaced with a security holding package on npm.
When metadata was still available, the projects showed thousands of dependents with a record of less than 10 weekly downloads.
This is a clear indicator for a sybil attack on teaRank.

\begin{figure}
    \alt{Horizontal bar visualizing the classification of sampled potentially sybil packages. Values: create-next-app (26), wallet/chains (16), no code (15), static strings (12), package clone (10), unpublished or private (8), security holding (4), other (9).}
    \centering
    \begin{tikzpicture}
      \def\total{100}
      \def\a{26}
      \def\b{16}
      \def\c{15}
      \def\d{12}
      \def\e{10}
      \def\f{8}
      \def\g{4}
      \def\h{9}
      
      \def\scale{8.2}

      \definecolor{colorA}{RGB}{66, 133, 244}
      \definecolor{colorB}{RGB}{219, 68, 55}
      \definecolor{colorC}{RGB}{204, 150, 0}
      \definecolor{colorD}{RGB}{15, 157, 88}
      \definecolor{colorE}{RGB}{171, 71, 188}
      \definecolor{colorF}{RGB}{0, 162, 180}
      \definecolor{colorG}{RGB}{233, 30, 99}
      \definecolor{colorH}{RGB}{158, 158, 158}

      \draw[fill=colorA] (0,0) rectangle ({\a/\scale},1) node[midway,white] {\small\bfseries \a};
      \draw[fill=colorB] ({\a/\scale},0) rectangle ({(\a+\b)/\scale},1) node[midway,white] {\small\bfseries \b};
      \draw[fill=colorC] ({(\a+\b)/\scale},0) rectangle ({(\a+\b+\c)/\scale},1) node[midway,white] {\small\bfseries \c};
      \draw[fill=colorD] ({(\a+\b+\c)/\scale},0) rectangle ({(\a+\b+\c+\d)/\scale},1) node[midway,white] {\small\bfseries \d};
      \draw[fill=colorE] ({(\a+\b+\c+\d)/\scale},0) rectangle ({(\a+\b+\c+\d+\e)/\scale},1) node[midway,white] {\small\bfseries \e};
      \draw[fill=colorF] ({(\a+\b+\c+\d+\e)/\scale},0) rectangle ({(\a+\b+\c+\d+\e+\f)/\scale},1) node[midway,white] {\small\bfseries \f};
      \draw[fill=colorG] ({(\a+\b+\c+\d+\e+\f)/\scale},0) rectangle ({(\a+\b+\c+\d+\e+\f+\g)/\scale},1) node[midway,white] {\small\bfseries \g};
      \draw[fill=colorH] ({(\a+\b+\c+\d+\e+\f+\g)/\scale},0) rectangle ({(\a+\b+\c+\d+\e+\f+\g+\h)/\scale},1) node[midway,white] {\small\bfseries \h};

      
      \node[anchor=north] at ({\a/2/\scale}, -0.2) {\scriptsize create-next-app};
      \node[anchor=south] at ({(\a+\b/2)/\scale}, 1.2) {\scriptsize Wallet/Chains};
      \node[anchor=north] at ({(\a+\b+\c/2)/\scale}, -0.2) {\scriptsize No Code};
      \node[anchor=south, text width=25pt] at ({(\a+\b+\c+\d/2)/\scale}, 1.2) {\centering\scriptsize Static\\Strings};
      \node[anchor=north, text width=29pt] at ({(\a+\b+\c+\d+\e/2)/\scale}, -0.2) {\centering\scriptsize Package\\\hspace*{5pt}Clone};
      \node[anchor=south, text width=44pt] at ({(\a+\b+\c+\d+\e+\f/2)/\scale}, 1.2) {\centering\scriptsize Unpublished\\\hspace*{5pt}or Private};
      \node[anchor=north, text width=29pt] at ({(\a+\b+\c+\d+\e+\f+\g/2)/\scale}, -0.2) {\centering\scriptsize Security\\Holding};
      \node[anchor=south] at ({(\a+\b+\c+\d+\e+\f+\g+\h/2)/\scale}, 1.2) {\scriptsize Other};
    \end{tikzpicture}
    \caption{Classes of sybil packages in a sample of 100 packages}
    \label{fig:sybil-sample}
\end{figure}
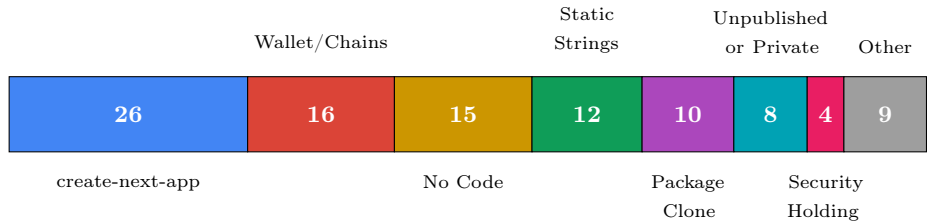

To analyze this phenomenon, we examine package metadata from npm.
As we try to approximate the order of magnitude for this attack, we employ the following heuristic-based methodology:
Initially, all npm packages registered in the tea testnet are classified as sybil if they are published after 2024, have less than 10 published versions and fulfill one of the following three criteria:
\begin{itemize}
  \item More than 95\% of transitive dependencies must have been created within 4 weeks of the creation of the package under consideration.
  \item More than 80\% of dependents must have more than 100 dependencies.
  \item The package was unpublished or marked as security holding.
\end{itemize}
In addition, we consider all transitive dependents of a sybil package to be sybil.
This methodology is conservative and it can be assumed to be unlikely that a legitimate package is flagged as sybil.
Overall, this results in the detection of 71,710 sybil packages on npm (2\% of all publicly listed packages on npm).
To confirm the reliability of the heuristic, we sample 100 potentially sybil packages uniformly at random and manually classify them.
We do not find any erroneous classification.
By calculating the one-tailed confidence interval for sample proportions, we can be 95\% certain that the population contains at most 3\% false positives.
The sample reveals multiple classes of automatically generated packages (see \Cref{fig:sybil-sample}).
Partially, the generation scripts can still be found in the artifacts of the packages.
Most commonly, boilerplate created by \texttt{create-next-app}, a bogus library detectable through \texttt{wallet.js} and \texttt{chains.js}, an artifact without JavaScript inside or a function returning a string are found.
Comparing these with unpublished packages, we assume that these contents prevailed against npm's spam detection.

These sybil attacks do not only increase the teaRank of projects but also affect other impact metrics.
Past research has considered the top $N\leq 1000$ most depended upon packages as benign packages for evaluating malware protection~\cite{Ferreira2021,Pohl2024}, evaluating the adoption of security best practices~\cite{Kabir2022} and others~\cite{Chao2019}.
We find that, at the time of writing, 532 of the top 1000 most directly depended upon packages are in our sybil set.
Though when taking transitive relationships into account, no package of the top 1000 most depended upon is marked as sybil by our approach.
As a significant fraction of sybil packages was unpublished, the actual figures were likely to be higher in mid 2024.
Since tea was introduced at the start of 2024, it is unlikely that the results of the referenced papers are affected by these sybil attacks.
Still, this raises questions on the validity of using this or similar impact metrics for software package focused research.

\section{Increasing Resilience against Unreliable Metrics}
\label{sec:resilience}

After identifying the largest risks for fraud and spoofing of popularity metrics.
Overall, we deem manual qualitative assessment as most resilient method for determining a project's criticality.
In total, we suggest 3 steps to minimize chances of being susceptible to spoofing.
\begin{enumerate}
  \item Historic developments of impact metrics require the highest effort for spoofing and should be part of the analysis.
  \item Maintain a high awareness for inconsistencies, e.g., many dependents but few downloads.
  \item Acquire a set of trust anchors, i.e., confirmed benign projects, and try to obtain as many (transitive) references from the trust anchors to the applicant.
\end{enumerate}

If assessment needs to be quantitative and largely automatic, there is an inherent persisting risk of being susceptible to spoofing.
Still, there are ways with the potential to reduce this risk.
On the one hand, a distinct impact metric could be designed.
All considered impact metrics are, at most, tied to digital identities which can be arbitrarily created.
In contrast, binding an impact metric to physical identities does not fully fix the reliability but drastically increases the effort and cost required for spoofing.
On the other hand, if the impact metrics are used for quantitative studies, such as for software engineering or security research, then a manual sampling process can be beneficial.
\Cref{sec:dependency-relations} has shown how, even if the ground truth is unknown, a fix-sized sample can greatly increase the trust of the population indeed representing the desired properties.

\begin{mdframed}
Answering \textbf{RQ1}, we find evidence for spoofing of all quantitative metrics and conclude that the metrics are rather unreliable.
Still, there are options to increase the resilience in the presence of metric spoofing and, depending on the concrete use case, there are options to employ these metrics in a valid way.
\end{mdframed}

\section{Related Work}
\label{sec:related-work}

\Citeauthor{Mujahid2022}~\cite{Mujahid2022} observe limitations of current popularity metrics such as stars and downloads and instead suggest employing package centrality.
They successfully apply this approach to identify npm packages in decline which is a criterion to prefer alternative dependencies.
\Citeauthor{Coelho2020}~\cite{Coelho2020} present a machine learning model as a metric for measuring the maintenance status of GitHub software projects.
As the input features are covered by the atomic metrics analyzed in this work, it is susceptible to the same degree of spoofing.
In \Cref{sec:metric-reliability}, the CHAOSS project popularity metric is considered.
Besides that, \citeauthor{CHAOSS2025}~\cite{CHAOSS2025} recently initiated a working group to develop methods for measuring funding impact.
Similarly, \citeauthor{Osborne2024b}~\cite{Osborne2024b} develops a toolkit for measuring the impact of public funding on \gls{foss}.
They argue that quantitative data has a \textquote{risk of creating perverse incentives through metric selection/optimisation} and suggests qualitative and mixed-methods to capture social, economic and technological impact.
This is in line with our results that assessment based on quantitative impact metrics is highly susceptible to fraud.

\section{Conclusion}
\label{sec:conclusion}

In this study, we explored the reliability of popularity and impact metrics for \gls{foss} project assessment.
Through dissecting existing measures into their atomic metrics, we create a classification into 7 distinct metric categories.
We find evidence that metrics from the 6 quantatively measured categories can be spoofed with at most medium effort.
For the spoofability of dependency relations, this involved the revelation of a large-scale sybil attack on npm for a campaign aimed at spoofing the teaRank metric.
To combat the impact of spoofing, we propose manual assessment as primary countermeasure since it is very challenging for a fraudster to paint a complete picture.
If the scale is too large for qualitative assessments, quality assurance mechanisms such as manually analysis of a uniform sample reduces risks.

We identify two major directions for future work based on these results.
First, a compilation of quantitative data measured in regular intervals could be utilized to detect sybil attacks on package repositories.
While, at first, this may sound contradictory to our result that all considered atomic metrics can be spoofed, we do not expect this to happen for no reason.
Sybil attacks on repositories as observed in this work are usually targeted and only focus on the subset of relevant metrics.
This enhances the probability of detection by an untargeted approach.
Second, package repositories are popular data sources for academic research due to their size and (semi-)structured data.
With the occurrence of sybil attacks on this dataset, the results' robustness in the presence of an attack is a novel and relevant research direction.
For example, it is intuitively unclear how significant the performance of malware detection systems on npm degrades if they are partially trained on bogus packages.

\printbibliography

\end{document}